\documentclass[
aps,nature,showpacs,superscriptaddress,numerical,amsmath,amssymb,floatfix,longbibliography,
reprint,
]{revtex4-1}
\usepackage[usenames,dvipsnames]{xcolor}
\usepackage[
pdffitwindow=true,
colorlinks=true,
frenchlinks=false,
linkcolor=blue,
anchorcolor=blue,
citecolor=blue,
filecolor=blue,
urlcolor=blue,
bookmarks=true,
bookmarksopen=true,
bookmarksnumbered=true,
bookmarksopenlevel=1,
plainpages=false,
pdfpagelayout=TwoPageLeft,
pdfpagelabels=true,
breaklinks
]{hyperref}
\usepackage[per-mode=symbol,separate-uncertainty]{siunitx}
\usepackage{graphicx}
\usepackage{dcolumn}
\usepackage{bm}
\usepackage[english]{babel}
\usepackage{color}
\usepackage{ulem}
\usepackage{nicefrac}

\begin{document}
	\title[]{Observation of Antiferromagnetic Magnon Pseudospin Dynamics and the Hanle effect}

	\author{T.~Wimmer}
	\email[]{tobias.wimmer@wmi.badw.de}
	\affiliation{Walther-Mei{\ss}ner-Institut, Bayerische Akademie der Wissenschaften, 85748 Garching, Germany}
	\affiliation{Physik-Department, Technische Universit\"{a}t M\"{u}nchen, 85748 Garching, Germany}
	\author{A.~Kamra}
	\affiliation{Center for Quantum Spintronics, Department of Physics, Norwegian University of Science and Technology, NO-7491 Trondheim, Norway}
	\author{J.~G\"uckelhorn}
	\affiliation{Walther-Mei{\ss}ner-Institut, Bayerische Akademie der Wissenschaften, 85748 Garching, Germany}
	\affiliation{Physik-Department, Technische Universit\"{a}t M\"{u}nchen, 85748 Garching, Germany}
	\author{M.~Opel}
	\affiliation{Walther-Mei{\ss}ner-Institut, Bayerische Akademie der Wissenschaften, 85748 Garching, Germany}
	\author{S.~Gepr{\"a}gs}
	\affiliation{Walther-Mei{\ss}ner-Institut, Bayerische Akademie der Wissenschaften, 85748 Garching, Germany}
	\author{R.~Gross}
	\affiliation{Walther-Mei{\ss}ner-Institut, Bayerische Akademie der Wissenschaften, 85748 Garching, Germany}
	\affiliation{Physik-Department, Technische Universit\"{a}t M\"{u}nchen, 85748 Garching, Germany}
	\affiliation{Munich Center for Quantum Science and Technology (MCQST), Schellingstr. 4, D-80799 M\"{u}nchen, Germany}
	\author{H.~Huebl}
	\affiliation{Walther-Mei{\ss}ner-Institut, Bayerische Akademie der Wissenschaften, 85748 Garching, Germany}
	\affiliation{Physik-Department, Technische Universit\"{a}t M\"{u}nchen, 85748 Garching, Germany}
	\affiliation{Munich Center for Quantum Science and Technology (MCQST), Schellingstr. 4, D-80799 M\"{u}nchen, Germany}
	\author{M.~Althammer}
	\email[]{matthias.althammer@wmi.badw.de}
	\affiliation{Walther-Mei{\ss}ner-Institut, Bayerische Akademie der Wissenschaften, 85748 Garching, Germany}
	\affiliation{Physik-Department, Technische Universit\"{a}t M\"{u}nchen, 85748 Garching, Germany}
	\date{\today}

	
	\begin{abstract}
		We report on experiments demonstrating coherent control of magnon spin transport and pseudospin dynamics in a thin film of the antiferromagnetic insulator hematite utilizing two Pt strips for all-electrical magnon injection and detection. The measured magnon spin signal at the detector reveals an oscillation of its polarity as a function of the externally applied magnetic field. We quantitatively explain our experiments in terms of diffusive magnon transport and a coherent precession of the magnon pseudospin caused by the easy-plane anisotropy and the Dzyaloshinskii-Moriya interaction. This experimental observation can be viewed as the magnonic analogue of the electronic Hanle effect and the Datta-Das transistor, unlocking the high potential of antiferromagnetic magnonics towards the realization of rich electronics-inspired phenomena.
	\end{abstract}

	\maketitle

	\begin{figure*}[ht!]%
		\includegraphics[width=\textwidth]{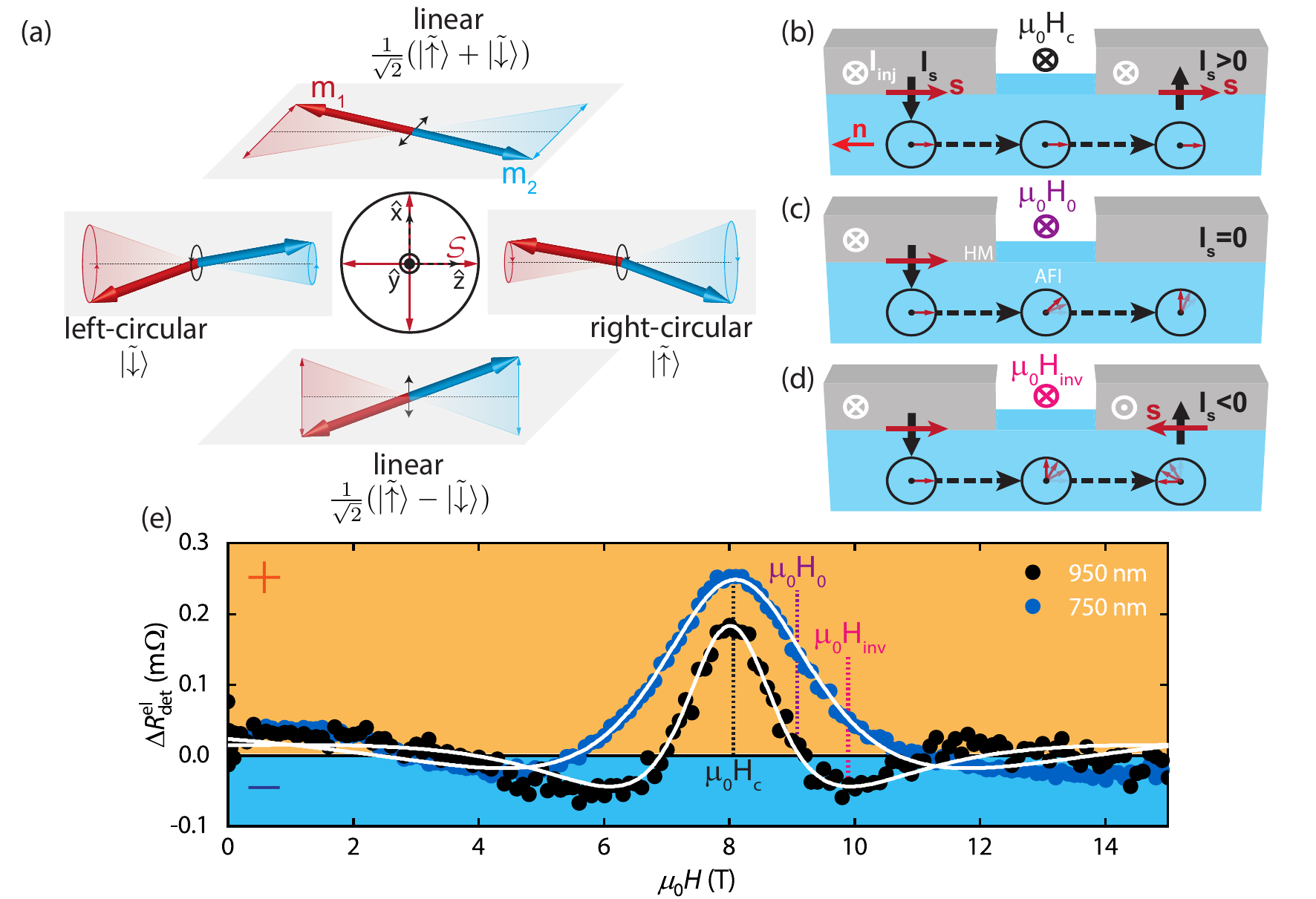}%
		\caption{(a) Pseudospin $\bm{\mathcal{S}}$ description of magnonic excitations obtained by linear superpositions of spin-up and \mbox{-down} antiferromagnetic magnons that correspond to right- and left-circular precessions of the N\'eel vector $\bm{n}$, respectively. A pseudospin collinear with the $z$-axis corresponds to spin-up or -down magnons carrying spin $\pm 1$. As the pseudospin rotates away from the $z$-axis, the precession of the N\'eel vector becomes increasingly elliptical merging into a linear oscillation for $\bm{\mathcal{S}} \parallel \hat{\bm{x}}$, corresponding to zero-spin excitations. The $z$-component of pseudospin $\mathcal{S}_z$ determines the actual magnonic spin which is probed in our measurements. (b), (c) and (d) Magnonic spin along $\hat{\bm{z}}\parallel \bm{n}$ is injected and detected respectively by the left and right heavy metal (HM) electrodes deposited on an antiferromagnetic insulator (AFI). The pseudospin precesses with a frequency controlled by the applied magnetic field while diffusing from the injector to the detector. As a result, positive (b), zero (c), or negative (d) magnon spin is detected giving rise to an analogous behavior of the measured spin signal between the two electrodes as shown in (e). The white curves depict the theoretical model fit (Eq.~\eqref{eq:fit}) to the experimental data shown via black and blue circles for devices with an injector-detector distance of $d=\SI{950}{\nano\metre}$ and $d=\SI{750}{\nano\metre}$, respectively, at $T=\SI{200}{\kelvin}$.}
		\label{fig:scheme}%
	\end{figure*}

	The different phases of electronic matter manifesting distinct transport properties are cornerstones of condensed matter physics and modern technologies. The electron spin together with spin-orbit interaction plays a fundamental role in hosting and controlling several of these phases, such as topological insulators~\cite{Hasan2010,Tokura2019}. Spin-dependent electronic transport has further underpinned industrial devices such as magnetoresistive read heads and memories. In these spin-electronic phenomena, {the} spin-orbit interaction results in an incoherent loss of spin currents, but can also be exploited for coherent control of spin and its transport~\cite{Datta1990,Manchon2019}.
	
	An emerging paradigm for spin and information transport via magnons in magnetic insulators offers distinct advantages~\cite{Bauer2012,Chumak2015,CornelissenMMR,Nakata2017,Yuan2018,Althammer2018,Klaui2018,Hou2019,Wimmer2019,Kamra2019}. While ferromagnetic magnons carry spin in only one direction, antiferromagnetic magnons come in pairs with opposite spins or N\'eel order precession chiralities. The latter can combine to form zero-spin excitations corresponding to linearly polarized oscillations of the N\'eel order~\cite{Kamra2017,Liensberger2019}. In general, the pairs of antiferromagnetic magnons and their superpositions can be described via a pseudospin~\cite{Cheng2016,Daniels2018,Shen2020,Kawano2019} in a manner similar to the actual spin of an electron (Fig.~\ref{fig:scheme}(a)). Besides the unique magnonic pseudospin feature, antiferromagnets also offer crucial advantages such as immunity to stray fields~\cite{Jungwirth2016,Baltz2018}, THz magnon frequencies~\cite{Jungwirth2016,Baltz2018,Li2020,Vaidya2020}, and ultrafast response times~\cite{Nmec2018,Olejnk2018}. Within our chosen convention, the $z$-component of such a pseudospin corresponds to the measurable magnon spin, while the transverse component characterizes the mode ellipticity and corresponds to zero-spin excitations. The formal equivalence between electron spin and antiferromagnetic magnon pseudospin has been predicted to result in a range of phenomena that are completely analogous in electronic systems and antiferromagnetic insulators (AFIs)~\cite{Cheng2016,Cheng2016B,Zyuzin2016,Daniels2018,Kawano2019,Kawano2019B,Shen2020}. The {experimental realizations of these theoretical predictions} promise to lift antiferromagnetic magnonics to a new level of functionalities. Here, we report the first observation of the magnonic analogue of the electronic Hanle effect~\cite{Fabian2007,Kikkawa1999,Jedema2002}. This is achieved by realizing the coherent control of {the} magnon spin and transport in a thin AFI.
	
	 In our experiments, spin current is injected from a heavy metal (HM) strip into an adjacent AFI via the spin Hall effect (SHE), producing an excess of spin-up magnons~\cite{Takei2014,Sinova2015,Klaui2018}. The injection thus creates a magnon pseudospin density directed along $\hat{\bm{z}}$ (Fig.~\ref{fig:scheme}(b)-(d)). In the presence of an easy-plane anisotropy and Dzyaloshinskii-Moriya interaction (DMI), spin-up and \mbox{-down} magnons are coherently coupled and therefore no longer eigenexcitations~\cite{Kawano2019,Rezende2019}. As a result, the pseudospin precesses in the $x$-$z$ plane with time while the magnons diffuse away from the injector. Its precession frequency $\Omega$ is determined by the anisotropy and a combination of the DMI field and canting-induced net magnetic moment. We control the latter by an external magnetic field and hereby obtain a handle on $\Omega$. At the compensation field $H_\mathrm{c}$, the anisotropy and the DMI contributions just cancel, resulting in $\Omega = 0$. The pseudospin, in this case, propagates through the AFI without any precession (Fig.~\ref{fig:scheme}(b)). In contrast, for the field $H_{\mathrm{0}}$, the pseudospin of the magnons arriving at the detector electrode points orthogonal to the $z$-axis (Fig.~\ref{fig:scheme}(c)). This corresponds to a linearly polarized pseudospin configuration with zero magnon spin density and thus a vanishing magnon spin signal at the detector (Fig.~\ref{fig:scheme}(e)). For $H_{\mathrm{inv}}$, the magnon pseudospin and actual spin densities have reversed directions while propagating from injector to detector (Fig.~\ref{fig:scheme}(d)). This situation corresponds to a negative magnon spin signal observed in our experiments (Fig.~\ref{fig:scheme}(e)).
	
		\begin{figure}[]%
		\includegraphics[]{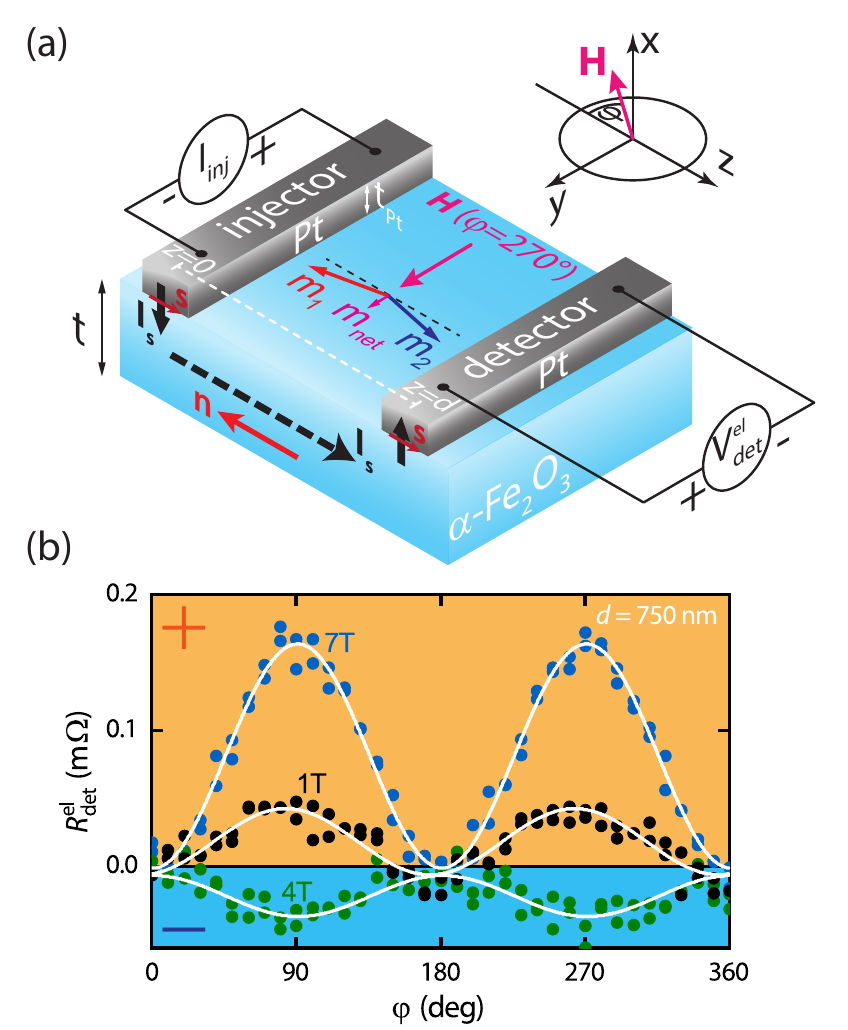}%
		\caption{(a) Sketch of the device geometry, the electrical wiring and the coordinate system. The canting of the magnetic sublattices $\bm{m}_1$ and $\bm{m}_2$ and the corresponding net moment $\bm{m}_\mathrm{net}$ as well as the N\'eel order parameter $\bm{n}$ are illustrated. Upon applying a charge current $I_\mathrm{inj}$ to the injector, a spin current $I_\mathrm{s}$ with spin polarization $s$ is generated via the SHE and injected into the hematite ($\alpha$-Fe$_2$O$_3$) with thickness $t$. The emerging antiferromagnetic magnon current is then detected via the inverse SHE-induced current at the detector by measuring the electrical voltage drop $V_\mathrm{det}^\mathrm{el}$. (b) Angle dependent magnon spin signals $R^\mathrm{el}_\mathrm{det}\propto V_\mathrm{det}^\mathrm{el}/I_\mathrm{inj}$ for electrically excited magnons measured at the detector for $T=\SI{200}{\kelvin}$ with a center-to-center distance of $d=\SI{750}{\nano\metre}$. The white solid lines are fits to a $\sin^2(\varphi)$-type function.
		}%
		\label{fig:ADMR}%
	\end{figure}

	We employ a $t = \SI{15}{\nano\metre}$ thin film of hematite ($\alpha$-Fe$_2$O$_3$) as the AFI. Our film is characterized by an easy $y$-$z$-plane anisotropy and an out-of-plane DMI vector. The thin hematite layer features an easy-plane phase over the entire temperature range and therefore lacks the Morin transition~\cite{Morin1950} (see SI for details~\footnotemark[1]), consistent with similar films~\cite{Han2020}. The equilibrium N\'eel vector $\bm{n}$ and the sublattice magnetizations $\bm{m}_{1,2}$ thus lie in the $y$-$z$-plane with a small canting angle between $\bm{m}_1$ and $\bm{m}_2$ (Fig.~\ref{fig:ADMR}(a)). An applied magnetic field along $\hat{\bm{y}}$ orients the N\'eel vector along $- \hat{\bm{z}}$. The magnitude of the external magnetic field $\mu_0 H$ further controls the canting angle and the net induced magnet moment $\bm{m}_{\mathrm{net}} = \bm{m}_1 + \bm{m}_2$, both bearing a constant DMI-induced offset and a variable contribution linear in $\mu_0 H$. We use $t_\mathrm{Pt} = \SI{5}{\nano\metre}$ thick, sputtered platinum as the HM for electrically injecting and detecting magnonic spin~\cite{CornelissenMMR}. A charge current $I_\mathrm{inj}$ is fed through the injector featuring typical current densities of $J_\mathrm{inj}\sim\SI{2e11}{\ampere\per\metre\squared}$. As a result, a $z$-polarized electron spin accumulation is generated at the interface with the AFI, leading to a $z$-polarized magnon spin and pseudospin current in the AFI (c.f.~Fig.~\ref{fig:ADMR}(a)). The reverse process enables the detection of the magnon spin in the AFI at its interface with the detector electrode, which is measured as a charge current/voltage. We extract the electrical signals from this SHE-based magnon injection/detection scheme using the current reversal method~\cite{SchlitzMMR,KathrinLogik} (see also SI~\footnotemark[1]).

		\begin{figure*}[t!]%
		\includegraphics[width=\textwidth]{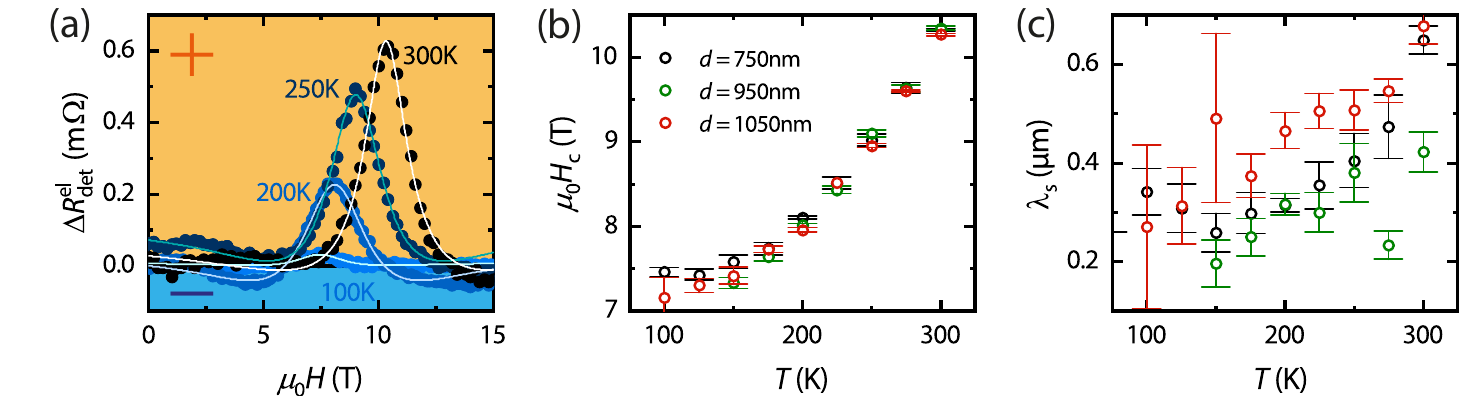}%
		\caption{(a) Electrically excited magnon spin signals $\Delta R^\mathrm{el}_\mathrm{det}\propto V_\mathrm{det}^\mathrm{el}/I_\mathrm{inj}$ for a structure with strip distance $d=\SI{750}{\nano\metre}$ plotted as a function of magnetic field for different temperatures. Light colored solid lines are fits to Eq.~\eqref{eq:fit}. (b) Compensation field $\mu_0 H_\mathrm{c}$ versus temperature extracted from experiments with devices of varying $d$. The temperature dependence of $\mu_0 H_\mathrm{c}$ follows the temperature trend of the uniaxial anisotropy of hematite. (c) Spin diffusion length $\lambda_\mathrm{s}$ as a function of temperature extracted from experimental data from different devices with varying $d$. $\lambda_\mathrm{s}$ increases with increasing temperature for all investigated structures.
		}%
		\label{fig:tempdep}%
	\end{figure*}
	
	For the configuration discussed above, the dynamics and diffusive transport of the magnon pseudospin density $\pmb{\mathcal{S}}$ in the AFI is described as~\cite{Kamra_theory}
	\begin{align}
	\frac{\partial \bm{\mathcal{S}}}{\partial t} & =  D \nabla^2 \bm{\mathcal{S}} - \frac{\bm{\mathcal{S}}}{\tau_\mathrm{s}} + \bm{\mathcal{S}} \times \Omega \, \hat{\mathbf{y}},
	\label{eq:spin_diffusion}
	\end{align}
	in direct analogy with the spin diffusion and dynamics for itinerant electrons~\cite{Fabian2007}. Here, $D$ is the magnon diffusion constant~\cite{CornelissenTheory} and $\tau_\mathrm{s}$ is the spin relaxation time accounting for the incoherent effect of spin-nonconserving interactions~\cite{Shen2020}. The pseudospin precession frequency $\Omega$ characterizes the coherent effect~\cite{Kamra2017,Liensberger2019,Shen2020} of spin-nonconserving, emergent spin-orbit~\cite{Kawano2019,Kawano2019B} interactions that couple spin-up and -down magnon modes. For $\Omega = 0$, Eq.~\eqref{eq:spin_diffusion} reduces to the magnon spin transport equation for easy-axis collinear AFIs~\cite{Shen2019,Troncoso2020}. In contrast, easy-plane anisotropy and canting-mediated noncollinearity in our AFI films break the rotational symmetry about the N\'eel order and coherently couple the opposite spin magnon modes. As detailed in the Supplementary Information (SI)~\footnote{\label{fn:supp}See Supplementary Information at [url], which includes Refs.~\cite{Nathans1964,Marmeggi1977,Eaton1969,Klaui2020,Lefever}, for details on growth conditions and magnetometry measurements of our hematite films, details on the measurement technique and geometry of the nanostructures, a rigorous description of the fitting routine of the field-dependent magnon spin signals, a more detailed study of the temperature-dependent magnon spin signals, a verification of the linearity of the current-voltage characteristics of the magnon spin signals as well as measurements ruling out electric leakage currents between the injector and the detector electrode. Additionally, we present a detailed description of the diffusive spin transport theory including the antiferromagnetic pseudospin dynamics.}, the resulting $\Omega$ is given by
	\begin{align}\label{eq:Omega}
	\hbar \Omega & = \hbar \omega_{\mathrm{an}} - \mu_0 H_{\mathrm{DMI}} m_{\mathrm{net}}  = \hbar \tilde{\omega}_{\mathrm{an}} - \mu_0 \tilde{m} H,
	\end{align}
	where $\tilde{\omega}_{\mathrm{an}}$ is a normalized anisotropy frequency and $H_{\mathrm{DMI}}$ is the effective DMI field.~$\tilde{m}$ is an equivalent magnetic moment that parametrizes the DMI strength~\cite{Morrish1995}. It allows for elucidating the linear $\mu_0 H$-dependence of the noncollinearity-mediated contribution to $\Omega$~\footnote{In general, $\Omega$ is expected to bear a constant contribution from anisotropy and a canting-mediated contribution that depends on the applied field $\mu_0 H$. These will differ for different materials and crystal structures, and $\Omega$ {\it vs.~}$\mu_0 H$ may be considered a Taylor expansion.}. Considering $z$-polarized magnon spin and pseudospin current density $j_{s0}$ injected by the electrode at $z = 0$, the steady state solution (see SI~\footnotemark[1]) to Eq.~(\ref{eq:spin_diffusion}) yields for the magnon spin density $s(z) = \mathcal{S}_z(z)$:
	\begin{align}
	s(z) & = \frac{j_\mathrm{s0}\lambda_{s}}{D(a^2+b^2)}\mathrm{e}^{-\frac{a z}{\lambda_\mathrm{s}}}\left(a\cos{\frac{b z}{\lambda_\mathrm{s}}}-b\sin{\frac{b z}{\lambda_\mathrm{{s}}}}\right),
	\label{eq:fit}
	\end{align}
	where $a \equiv \sqrt{(1+\sqrt{1+\Omega^2\tau_\mathrm{s}^2})/2}$, $b \equiv\sqrt{(-1+\sqrt{1+\Omega^2\tau_\mathrm{s}^2})/2}$, and $\lambda_s = \sqrt{D\tau_s}$ is the spin diffusion length. Equation~\eqref{eq:fit} describes the magnon spin density at a distance $z$ from the injector. It is proportional to the magnon spin signal measured by the detector electrode at $z = d$. Together, Eqs.~\eqref{eq:Omega} and~\eqref{eq:fit} describe the key phenomenon reported here and form the basis for analyzing our experimental data. In Fig.~\ref{fig:scheme}(e), corresponding theoretical curves (white solid lines) are shown together with experimental data (black and blue data points) for two devices featuring different electrode spacings $d$. Consistent with our model, we see a pronounced peak in the positive magnon spin signal regime for both devices. This peak corresponds to the compensation field $\mu_0 H_\mathrm{c}$ for which $\Omega = 0$. Due to the vanishing pseudospin precession frequency at $\mu_0 H_\mathrm{c}$, the peak position is independent of the electrode spacing $d$. For increasing field strength, the spin signal decreases until it approaches zero signal at $\mu_0 H_0$, corresponding to a $\SI{90}{\degree}$ rotation of the pseudospin vector, i.e.~a linear polarization of the propagating magnon modes carrying zero spin. A sign inversion of the spin signal is evident when the field is further increased to $\mu_0 H_\mathrm{inv}$, corresponding to a full $180^\circ$ rotation of the pseudospin vector $\bm{\mathcal{S}}$ and therefore an inversion of the magnon mode chirality/spin (c.f.~Fig.~\ref{fig:scheme}(a)). Since both $\mu_0 H_0$ and $\mu_0 H_\mathrm{inv}$ correspond to a finite precession frequency $\Omega$, their values are expected to vary with the spacing $d$ between the injector and detector electrodes, in agreement with our experimental data in Fig.~\ref{fig:scheme}(e). As evident, the same behaviour is observed for decreasing field strength $\mu_0 H < \mu_0 H_\mathrm{c}$, corresponding to a pseudospin precession in the opposite sense.

	Subsequently, we measure the magnon spin signal $R^\mathrm{el}_\mathrm{det}\propto V_\mathrm{det}^\mathrm{el}/I_\mathrm{inj}$ at the detector (see SI for details~\footnotemark[1]) as a function of the external magnetic field orientation $\varphi$ within the $y$-$z$-plane as illustrated in Fig.~\ref{fig:ADMR}(a). The result is shown in Fig.~\ref{fig:ADMR}(b) for a center-to-center strip distance of $d = \SI{750}{\nano\metre}$. The data exhibit a $180^\circ$-symmetric modulation consistent with the SHE-mediated spin injection and detection of magnons~\cite{CornelissenMMR,SchlitzMMR}. This corresponds to a $\sin^2(\varphi)$ angular variation of the signal, which is the expected dependence for electrically induced magnon transport~\cite{CornelissenMMR,SchlitzMMR}. A possible spin Seebeck effect, in contrast, would yield a $\sin(\varphi)$ dependence~\cite{CornelissenMMR,Ganzhorn2017}. Hence, the angle dependence can be fitted with a simple $\Delta R^\mathrm{el}_\mathrm{det}\sin^2(\varphi)$ function, where $\Delta R^\mathrm{el}_\mathrm{det}$ represents the amplitude of the electrical magnon spin signal. The signal modulation is shifted by {$\sim \SI{90}{\degree}$} compared to similar measurements on ferrimagnetic materials~\cite{CornelissenMMR,SchlitzMMR,Shan2017}. This is due to the fact that the electrical magnon excitation is only active when $\bm{\mu_\mathrm{s}}\parallel\bm{\mathrm{n}}$, i.e.~for $\bm{\mathrm{H}}\perp\bm{\mathrm{n}}$ in our experiments. Thus, we can confirm that the excited magnons in our experiments originate from the antiferromagnetic N\'eel order consistent with previous experiments in AFIs~\cite{Klaui2018}. Most importantly, we indeed observe two sign inversions of $R^\mathrm{el}_\mathrm{det}$ in the investigated field range. While a positive signal is measured for $\mu_0 H=\SI{1}{\tesla}$ and $\SI{7}{\tesla}$, a negative signal ensues at $\SI{4}{\tesla}$. These measurements are further evidence for the rotation of the pseudospin vector via the coherent coupling $\Omega$ between the antiferromagnetic magnon modes described in the spin diffusion equation~\eqref{eq:spin_diffusion}.
	
	Last but not least, we extract the relevant magnon transport parameters from our data using the diffusive spin transport model given in Eq.~\eqref{eq:fit}. To this end, we carried out temperature-dependent measurements of the field-dependent magnon spin signals $\Delta R^\mathrm{el}_\mathrm{det}$, which are shown in Fig.~\ref{fig:tempdep}(a). Here, light colored solid lines correspond to fits to Eq.~\eqref{eq:fit}. For the fitting routine, we consider a finite (constant) offset signal, which is added to Eq.~\eqref{eq:fit}. Furthermore, the free fit parameters used were $j_\mathrm{s0}$, $\lambda_\mathrm{s}$, $\tau_\mathrm{s}$, $D$ and $\tilde{\omega}_\mathrm{an}$, whereas $\tilde{m}$ was fixed to the value of the net magnetic moment at zero magnetic field (see SI for details~\footnotemark[1]). For all investigated temperatures and devices with varying $d$ we obain excellent agreement between our experiments and the theoretical model, strongly supporting the validity of our theory. As evident from Fig.~\ref{fig:tempdep}(a), we observe a decrease of the peak amplitude at $\mu_0 H_\mathrm{c}$ with decreasing temperature, which is expected from the electrically excited magnon transport effect~\cite{SchlitzMMR,CornelissenTemp,ZhangAFM,Fontcuberta2019} (see also SI~\footnotemark[1]). Moreover, we find a clear decrease of the compensation field with decreasing temperature in Fig.~\ref{fig:tempdep}(a). For a quantitative treatment of this behaviour, we extract $\mu_0 H_\mathrm{c}$ for each temperature from the fits (via $\tilde{\omega}_\mathrm{an}$) and plot its temperature dependence in Fig.~\ref{fig:tempdep}(b). For each structure, we observe a constant behaviour in the temperature range from $\SI{100}{\kelvin}$ to $\SI{150}{\kelvin}$. A significant increase is evident for larger temperatures up to $\SI{300}{\kelvin}$. As evident from Eq.~\eqref{eq:Omega}, the compensation field can be expressed as $\mu_0 H_\mathrm{c} = \hbar \tilde{\omega}_\mathrm{an}\,(\tilde{m})^{-1}$. Therefore, $\mu_0 H_\mathrm{c}$ directly corresponds to the normalized anisotropy energy $\tilde{\omega}_\mathrm{an}$ of the hematite. We thus expect that  $\mu_0 H_\mathrm{c}$ follows the temperature dependence of the easy-plane anisotropy. This is supported by previous measurements of the temperature dependence of the anisotropy energy in hematite, which qualitatively agree with the temperature dependence of $\mu_0 H_\mathrm{c}$~\cite{Besser1967}. Hence, our results support the assumption that the coupling strength $\Omega$ defined in Eq.~\eqref{eq:Omega} is related to the easy-plane anisotropy in hematite. Finally, we calculate the magnon diffusion length $\lambda_\mathrm{s}$ using the extracted diffusion constant $D$ and the spin relaxation time $\tau_\mathrm{s}$ from our fits. The obtained temperature dependence of $\lambda_\mathrm{s}$ is shown in Fig.~\ref{fig:tempdep}(c). Overall, we find an increase of $\lambda_\mathrm{s}$ with increasing temperature for all studied injector-detector distances $d$. At room temperature, we extract $\lambda_{s} \approx \SI{0.5}{\micro\metre}$, which is in perfect agreement with recent reports measuring the spin diffusion length in the easy-plane phase of hematite thin films using distance-dependent measurements~\cite{Han2020,Klaui_2020}.

	As a key result, we have experimentally demonstrated the coherent control of spin currents and magnon pseudospin dynamics in antiferromagnetic insulators. This opens new avenues for antiferromagnetic magnonic applications such as spin based transistors or field-controlled switchable devices. Moreover, our experimental exploitation of the magnonic equivalent of a {spin-1/2} electron system provides the first crucial step towards various pseudospin-based concepts such as an unconventional non-Abelian computing scheme~\cite{Daniels2018}.

	We gratefully acknowledge financial support from the Deutsche Forschungsgemeinschaft (DFG, German Research Foundation) under Germany’s Excellence Strategy -- EXC-2111 -- 390814868 and project AL2110/2-1, and the Research Council of Norway through its Centers of Excellence funding scheme, project 262633, ``QuSpin''. A.K. thanks Siddhartha Omar for valuable discussions.


%

\end{document}